\documentclass[iop]{emulateapj}
\shorttitle{A Hypervelocity Globular Cluster}
\shortauthors{Caldwell \etal~}
\def\etal{{\it et al.}}
\def\kms{\,km~s$^{-1}$}

\def\arcsec{\char'175 }

\def\hub{\ifmmode H_\circ\else H$_\circ$\fi}
\usepackage{subfigure}
\usepackage{amsmath}
\usepackage{graphics}

\begin{document}

\title{A Globular Cluster Toward M87 with a Radial Velocity $<-1000$~\kms: The First Hypervelocity Cluster}

\author{Nelson Caldwell\altaffilmark{1}, Jay Strader\altaffilmark{2}, Aaron J.~Romanowsky\altaffilmark{3,4}, Jean P.~Brodie\altaffilmark{4}, Ben Moore\altaffilmark{5}, Jurg Diemand\altaffilmark{5}, Davide Martizzi\altaffilmark{6}}

\email{caldwell@cfa.harvard.edu}

\altaffiltext{1}{Harvard-Smithsonian Center for Astrophysics, Cambridge, MA, 02138}
\altaffiltext{2}{Department of Physics and Astronomy, Michigan State University, East Lansing, MI 48824}
\altaffiltext{3}{Department of Physics and Astronomy, San Jos\'e State University, San Jose, CA 95192}
\altaffiltext{4}{University of California Observatories, Santa Cruz, CA 95064}
\altaffiltext{5}{Institute for Theoretical Physics, University of Zurich, CH-8057 Zurich, Switzerland}
\altaffiltext{6}{Department of Astronomy, University of California, Berkeley, CA 94720}

\begin{abstract}

We report the discovery of an object near M87 in the Virgo Cluster with an extraordinary blueshift of $-1025$ \kms, offset from the systemic velocity by $>2300$ \kms. Evaluation of photometric and spectroscopic data provides strong evidence that this object is a distant massive globular cluster, which we call HVGC-1 in analogy to Galactic hypervelocity stars. We consider but disfavor more exotic interpretations, such as a system of stars bound to a recoiling black hole. The odds of observing an outlier as extreme as HVGC-1 in a virialized distribution of intracluster objects are small; it appears more likely that the cluster was (or is being) ejected from Virgo following a  three-body interaction. The nature of the interaction is unclear, and could involve either a subhalo or a binary supermassive black hole at the center of M87.

\end{abstract}


\section{Introduction}
Extreme astrophysical objects are occasionally found in large samples of data. While classifying a million galaxy images, the Galaxy Zoo project found Hanny's Voorwerp, an unusual cloud ionized by an active galactic nucleus \citep{Lintott}. In a study of blue horizontal branch stars in the Galactic halo selected from the Sloan Digital Sky Survey, \cite{brown2005} found hypervelocity stars that are thought to originate in three-body interactions with the supermassive black hole (SMBH) at the center of the Galaxy. In an extensive study of star clusters, planetary nebulae, and \ion{H}{2} regions in M31, \cite{caldwell10} found one star with the most negative velocity known ($-780$~\kms), a probable member of the Andromeda giant stream. 

Here, we report on an object found in a different large survey that has an even more extreme negative velocity: an apparent globular cluster (GC) toward the central Virgo Cluster galaxy M87.

\section{Observations}\label{sec:obs}
We have been collecting spectra of GC candidates in the Virgo Cluster for several years, using Keck/DEIMOS and LRIS and MMT/Hectospec \citep{roman,strader2011}. Those papers reported roughly 500 new confirmed GCs. 

Our more recent dataset, taken mostly with MMT/Hectospec from 2010--2013, contains more than 5000 separate observations of 2500 candidate GCs and ultra-compact dwarfs, covering a non-uniform area within 1\degr\ of M87 and 0.5\degr\ of M60.

Details of the survey will be presented elsewhere, but in brief, Hectospec with the 270 l mm$^{-1}$ grating provided spectra with a resolution of 5 \AA\ over the range 3700--9200 \AA\ \citep{fab}. The total exposure times were 1--4 hours per field. Data were extracted and wavelength calibrated from the two-dimensional images, and sky was subtracted using dedicated fibers. Multiple observations were coadded; about half of the objects were observed more than once. Velocities were measured through cross-correlation as described in Strader \etal~(2011a). We estimate that $\sim1800$ objects have secure velocities, though this number is not yet final.

Of these 1800, more than 1000 objects have measured velocities between 500 and 3000 \kms, with a clear median $\sim 1300$ \kms. These are all likely Virgo Cluster members. The remainder are Galactic stars ($\sim600$ objects) or background galaxies ($\sim100$ objects). Figure 1 shows a preliminary histogram of velocities in our survey, plus GCs from Strader \etal~(2011a), showing clear peaks associated with the foreground star and Virgo GC populations. The distribution of Virgo galaxies is also plotted. The survey target with a velocity $<-1000$ \kms\ is the subject of this paper.

\begin{figure}[ht]
\includegraphics[width=3.2in]{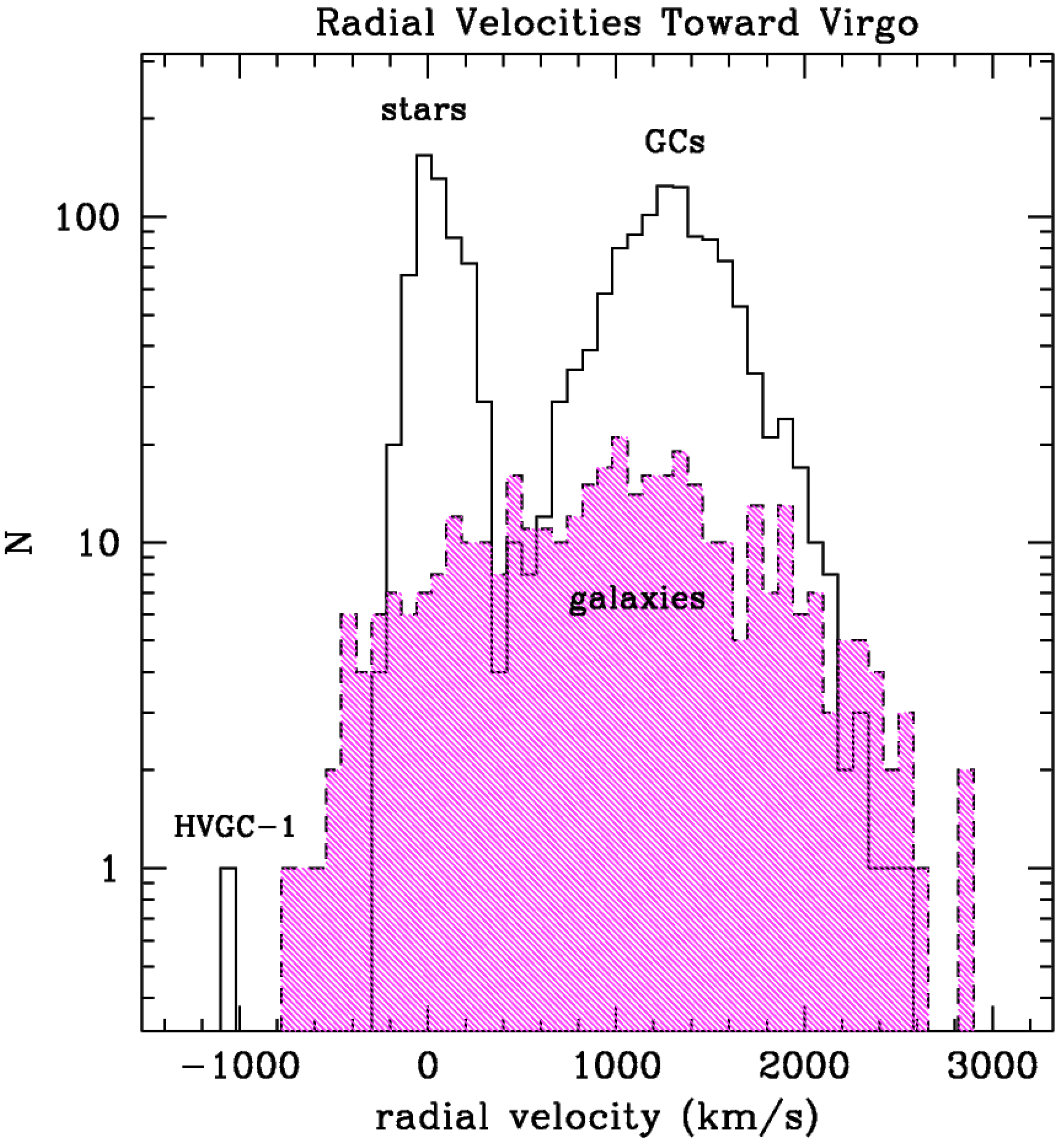}
\caption{Velocity distribution of objects toward Virgo, including all confirmed GCs, all Hectospec velocities, and galaxies (from Rines \& Geller 2008). The distinct stellar and GC distributions are clear, as is the broader galaxy distribution (dotted and shaded magenta). HVGC-1 is the marked extreme left outlier.}
\end{figure}

The J2000 decimal coordinates are (R.A.,Dec.) = (187.72791,+12.68295). It is located 17.6\arcmin\ north of M87, a projected distance of $\sim84$ kpc if the object has the same distance as M87 (we adopt 16.5 Mpc for consistency with Strader \etal~2011a). For reasons explained below, we dub the object HVGC-1 (for the first ``hypervelocity globular cluster"); in the nomenclature of Strader \etal~(2011a) for M87 GCs, this object has the catalog designation H70848.

We observed the object on three separate occasions, finding consistent results in each case. The final combined radial velocity is $-1026\pm13$ \kms. This is the most negative, bulk velocity ever measured for an astronomical object not orbiting another object.

\section{What is this object?}\label{sec:what}

The only reasonable possibilities are that this object is a GC in or near Virgo or an individual Galactic star. The extreme negative radial velocity is very difficult to explain if the source is a star, and only somewhat easier to explain if not. 


\subsection{A Star?}

The known Galactic hypervelocity stars all have positive velocities \citep{brown}, as expected if they have been ejected by the central SMBH into the halo. A highly negative velocity for an ejected star would be observed only in the unlikely event that a star was ejected towards the Sun, and quite recently ($\la8$ Myr, given a distance of $\sim8$ kpc). Our object has a galactic latitude of 74\degr, and thus is not in the direction of the Galactic Center. In our Hectospec sample, the next most negative velocities are $>-300$ \kms, above the expected escape velocity for Galactic halo stars \citep{kenyon}. Thus the other negative-velocity objects we observe are likely to be stars.

If the object were in a different part of the sky, we could take more seriously the exotic possibility that it might be a hypervelocity star from a nearby galaxy (such as M31; Sherwin \etal~2008). However, its velocity and position are very implausible for an origin from M31 or other nearby galaxies.

\subsection{A Star Cluster?}

If the object is in Virgo, consider that M87 has a systemic velocity of $1307\pm8$ \kms\ \citep{smith}, while the galaxy cluster itself has a mean of $1050\pm35$ \kms\ \citep{binggeli}, so that our object's velocity with respect to M87 and Virgo are about 2300 and 2100 \kms, respectively. 

The object could be confirmed as a GC if resolved with high-resolution optical data. For a first constraint on the half-light radius $r_h$, we used an archival $i$-band CFHT image of the field with seeing $\sim0.65$\arcsec. We measured $r_h$ for both HVGC-1 and, as a comparison, several luminous and large ($r_h\sim20$ pc) star clusters in the same field (Brodie \etal~2011). The estimates were made as discussed in Strader \etal~(2011a): using {\tt ishape} (Larsen 1999) we fit King models with fixed $c=30$ convolved with a point spread function made from bright stars near HVGC-1. The $r_h\sim20$ pc objects are clearly resolved and we reproduce the published $r_h$ estimates to within $\sim15$--20\%. HVGC-1 shows modest evidence for being resolved with $r_h\sim6$ pc. While we believe this measurement is too small to be reliable, the fitting---and the comparison to known objects---suggests an upper limit of $r_h\sim10$--15 pc. This is consistent with a GC, but rules out a larger, more distant galaxy infalling to Virgo.

We have obtained photometry of HVGC-1 using ground-based CFHT images taken for the Next Generation Virgo Cluster Survey (NGVS; Ferrarese \etal~2012) and the associated NGVS-IR $K$-band survey, but separately processed by CADC (Gwyn 2008). Foreground extinction corrections were taken from Schlafly \& Finkbeiner (2011). 

Mu\~{n}oz \etal~(2014) have found that in the optical/IR the best separation between single stars and composite old stellar populations is with $uiK$ photometry. Figure 2 shows a two-color $i-K$ vs.~$u-i$ diagram using CADC photometry of objects spectroscopically classified as stars or GCs and with photometric uncertainties $<0.05$ mag. Here we consider an object a foreground star if its Hectospec velocity is $<250$ \kms; objects with velocities between 500 and 2300 \kms\ are designated as GCs. As found in  Mu\~{n}oz \etal~(2014), the GC and star sequences cleanly separate, with only a few outliers. These may be objects with problematic photometry.

With $u-i=1.67\pm0.01$ and $i-K=-0.01\pm0.03$, HVGC-1 is firmly situated in the GC sequence, strongly suggesting that it is a GC rather than a single star. It has $g_0 = 20.57$, consistent with a massive GC if at the distance of M87 (see below).

\begin{figure}[ht]
\includegraphics[width=3.4in,angle=0]{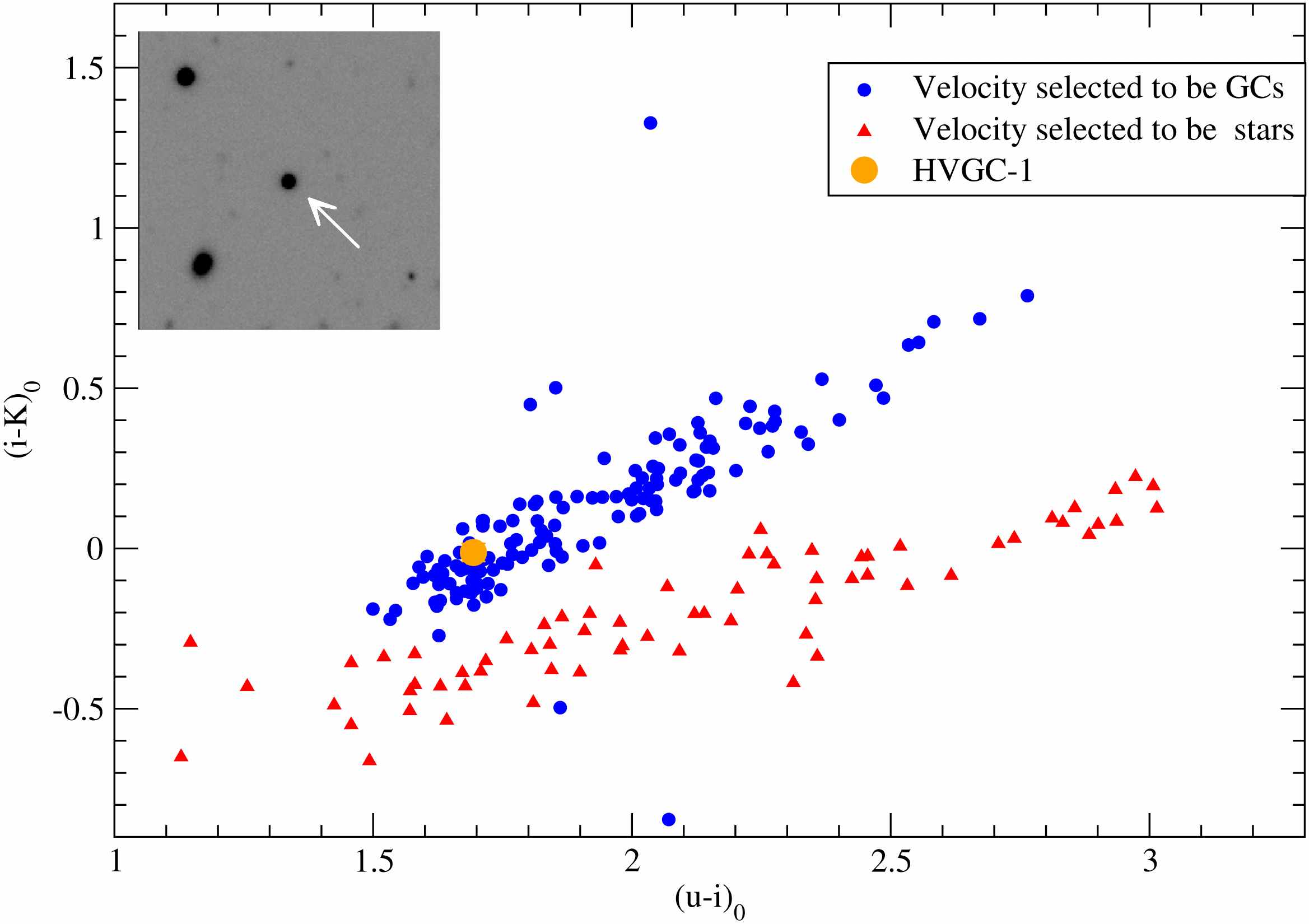}
\caption{A two-color $uiK$ diagram, derived for objects with Hectospec velocity classification. Objects with velocities between 500 and 2300 \kms\ are considered GCs; those $<250$ \kms\ stars. HVGC-1 clearly falls in the GC sequence. A CFHT $g$-band cutout (40\arcsec) is inset. \label{colors}}
\end{figure}

The Hectospec spectrum of HVGC-1 shows strong lines of \ion{Ca}{2} H\&K, moderate strength Balmer lines, and a weak G-band (Figure 3; right). There are no emission lines, other than night-sky line residuals. Overall the spectrum looks like an early G star or an intermediate-metallicity GC (Caldwell \etal~2011), not at all like the B and A HVS of \cite{brown}.

\cite{rose} discussed the use of line index ratios to search for young stars in composite stellar populations. These ratios can also be used to distinguish single stars from simple stellar populations. In Figure 3 (left) we plot the ratio of central intensities of \ion{Ca}{2} H+H$\epsilon$ to \ion{Ca}{2} K (the index ``CaII'') against the ratio of H$\gamma$ to the G-band at 4300 \AA, though this latter choice is not essential. 

In Figure 3 (as in Figure 2), we use the M87 candidates classified by velocity into star or GC categories. There is a clear separation between the line-index ratio sequences. Adding confirmed M31 GCs \citep{caldwell11} and Galactic halo stars observed with Hectospec strengthens this conclusion. Foreground stars are mostly in the upper part of the diagram, overlapping the more metal-rich M31 and Virgo GCs (these objects have few hot stars, so the CaII index offers no discrimination). For HVGC-1, the CaII ratio is around 0.5, and though the uncertainties are large, this ratio is lower than observed for any known foreground star in our sample. Rather, the index ratios are within the range for confirmed GCs in M87 and M31.

\begin{figure*}[ht]
\includegraphics[width=6.7in,angle=0]{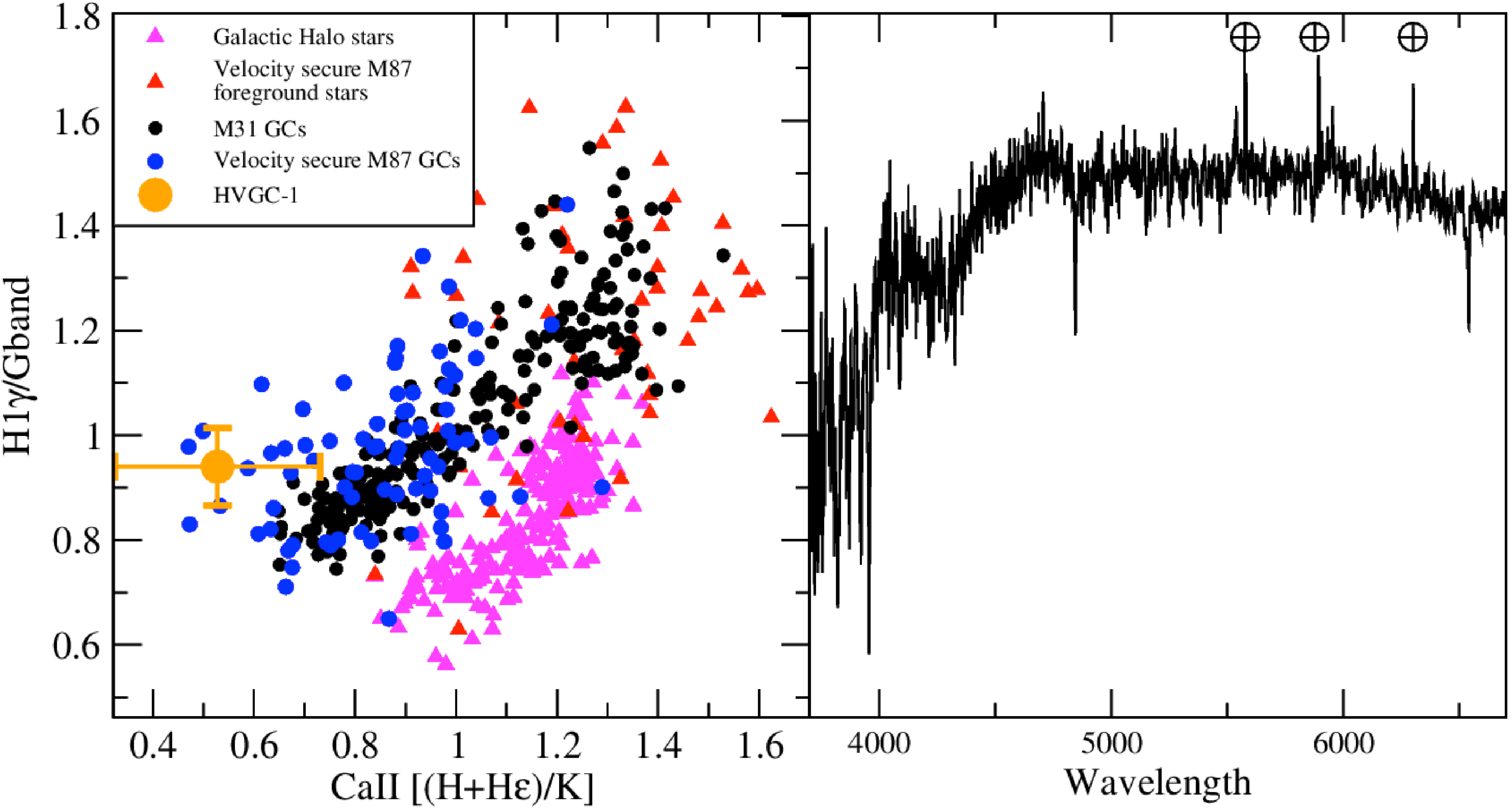}
\caption{Left: Line-index ratios for stars and GCs. The ratio of H$\gamma$/G-band is plotted vs.~CaII H+H$\epsilon$/K. Different symbols are probable GCs (those with velocities between 500 and 2300 \kms) and probable stars ($<250$ \kms). Also shown are other Galactic halo stars and GCs from M31. HVGC-1 has line indices more similar to GCs than individual stars. Right: The spectrum of HVGC-1, with residual sky lines marked.\label{indices}}
\end{figure*}

The combined photometric and spectroscopic data provide strong evidence that HVGC-1 is a GC.

Given this interpretation, we can calculate basic properties of the cluster. Using the calibration of Lick Fe indices from GC integrated light spectra with [Fe/H] provided in \cite{caldwell11}, we derive a spectroscopic [Fe/H] estimate of  [Fe/H]=--$0.9\pm0.3$. This value is consistent with the intermediate-metallicity appearance of the spectrum (moderate Balmer/metal lines), though the photometry in Figure 2 suggests a somewhat lower metallicity. From the photometry we estimate $V_0=20.33$, which would imply $M_V = -10.76$ and a mass of $\sim3.4\times10^{6} M_{\odot}$ for a distance of 16.5 Mpc and assuming $M/L_V = 2$ (Strader \etal~2011b). However, it is possible the cluster is somewhat closer than this distance (see \S 5.5), which would make it less massive. The expected $\sigma$ for a typical GC size (2.5 pc) would be only $\sigma=24$ \kms; we constrain the Hectospec value to be $\la 80$ \kms.

\section{How unusual is the velocity?}\label{sec:velocity}

We have concluded the object is likely to be a GC. How does its velocity compare with other stellar systems in Virgo? \cite{binggeli99} and \cite{kara2010} used kinematic catalogs to discuss negative-velocity galaxies in Virgo (Figure 1 shows the velocity distribution of Virgo galaxies from Rines \& Geller 2008). Considering these catalogs, there are $>60$ galaxies in Virgo with published negative velocities, the most prominent being NGC~4406 (M86, with radial velocity --258 \kms). However, some of the most negative values may be inaccurate. For example, VCC 846 has a published velocity in these catalogs --730 \kms, but the SDSS value is --510 \kms. Spectra from Keck/ESI (Forbes \etal~2011) support the SDSS value (S.~Penny \& D.~Forbes, private communication).

The only galaxy with a confirmed velocity below --600 \kms\ is the dwarf elliptical (dE) VCC 815, with --743 \kms. This galaxy is located about 14\arcmin\ (69 kpc) in projection from M86. VCC 815 is therefore likely associated with M86, and indeed the whole M86 subgroup may be merging with the central M87 subgroup, generating higher velocities for some members \citep{binggeli99}. Given that the GC system of M86 has $\sigma = 292$ \kms\ (Park \etal~2012), it is possible that M86 has some GCs with velocities near --$1000$ \kms, though none so low have yet been measured (the lowest in Park \etal~2012 is --$864\pm57$ \kms). However, since the GC under consideration is not near M86 (it is over a degree away, compared to just 18\arcmin\ from M87), and we see no other plausible M86 GCs in our sample, we believe an association with M86 is very unlikely.

In fact, the unusual velocity does not appear to be consistent with the tail of the GC velocity distribution of any individual Virgo galaxy, including M87. For M87 in particular the velocity dispersion at this projected radius is about 300 \kms\ (Strader \etal~2011a), so HVGC-1 would be a more than $7\sigma$ outlier. We must look for other mechanisms to explain its relative velocity of over 2300 \kms\ with respect to M87.

\begin{figure}[ht]
\includegraphics[width=3.3in]{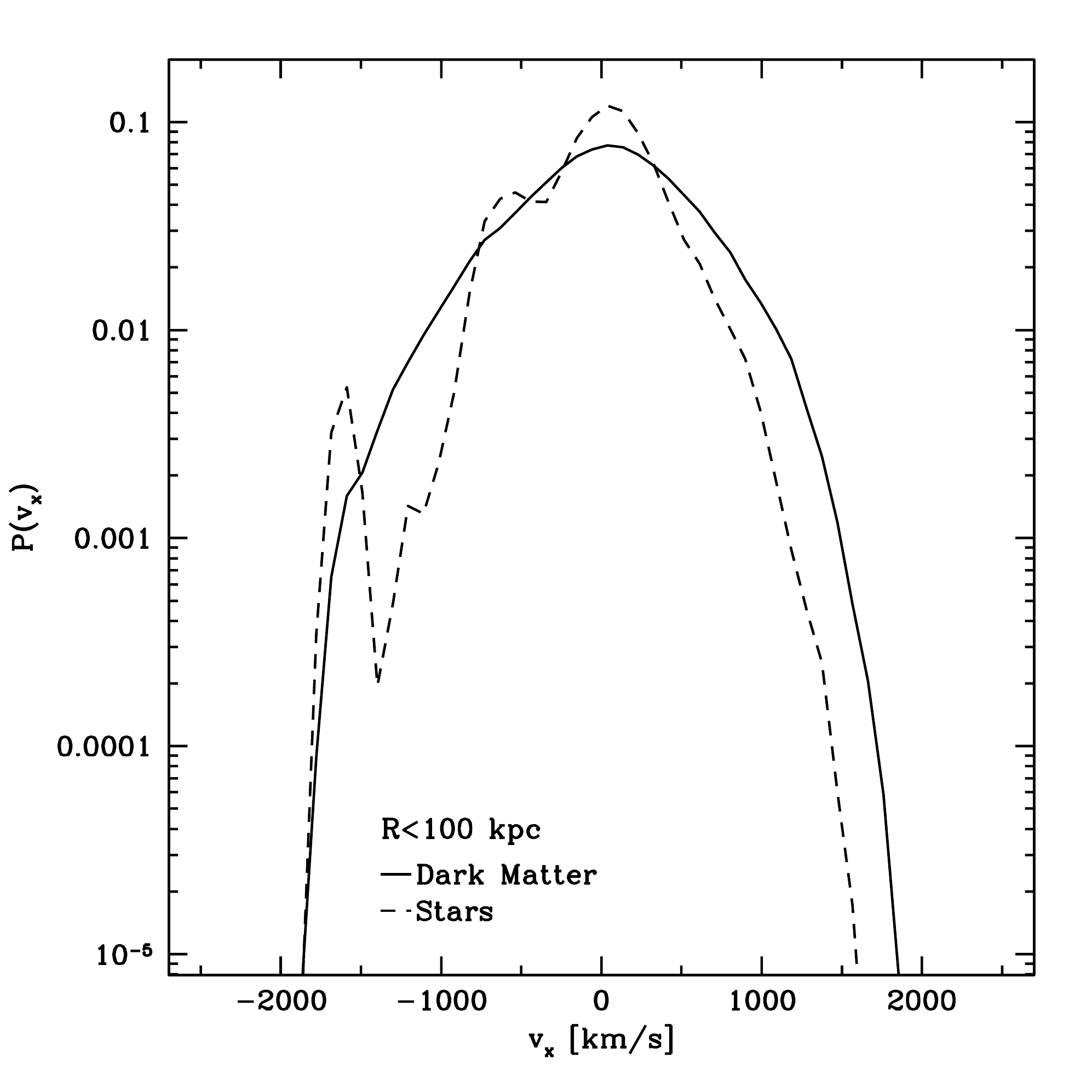}
\caption{Velocity distribution of stars and dark matter from the simulation of a Virgo-like cluster by Martizzi \etal~(2012). The plotted distribution is
(arbitrarily) chosen to be the projected X plane, considering all particles projected within 100 kpc of the cluster center. No particles with the relative velocity of HVGC-1 ($<-2300$ \kms) are found.
\label{vel4}}
\end{figure}

\section{The Origin and Future of  HVGC-1}

\subsection{The Virialized Intracluster Light}

All galaxy clusters have an important stellar component in the form of ``intracluster light": stars stripped from the outer parts of galaxies during the mergers or encounters that occur during the assembly of cluster-central galaxies like M87 (e.g., Purcell \etal~2007). Given their extended spatial distribution, GCs are expected to be stripped along with field stars, and have been found in intergalactic space in clusters including Virgo (Lee \etal~2010). Here we consider whether a GC with a relative velocity of  $-2300$ \kms\ is consistent with the tail of a virialized distribution of objects formed during the assembly of Virgo.

A simple argument that HVGC-1 is unlikely to be in the tail of a distribution is that there are no objects at less extreme outlying velocities: the most extreme positive velocity is $<2800$ \kms,  less than 1500 \kms\ offset from the M87 systemic, and there are no other other non-stellar objects with velocities $<-300$ \kms, 1600 \kms\ from systemic. Such objects should be \emph{much} more common than GCs like HVGC-1 if one is observing the tail of a virialized distribution.

As another approach, we use a simulation of the formation of a Virgo-like cluster ($M_{vir}\sim10^{14} M_{\odot}$) , which includes a central ``brightest cluster galaxy" (Martizzi \etal~2012). In Figure 4 we plot the model one-dimensional velocity distribution for both stars and dark matter projected within a 100 kpc radius cylinder around the cluster center. The stellar and dark matter distributions are close to Gaussian, with some lumps in the stellar distribution due to individual galaxies. Both the stars and dark matter have sharp cutoffs well short of a $>2300$ \kms\ relative velocity, with literally zero particles in the simulation at such an extreme velocity. This comparison should not be over-interpreted, as the simulation is not a perfect match to Virgo, but it does suggest that HVGC-1 is unlikely to be part of a virialized distribution of intracluster GCs.

\subsection{A Subhalo Interaction}

While HVGC-1 is probably not a normal intracluster GC, it is possible that it was recently given a ``kick" through a three-body interaction with M87 and a subhalo. Velocity outliers are sometimes observed in simulations of galaxy formation due to such interactions. For example, Sales \etal~(2007) suggested that the extreme radial velocity of the Galactic satellite Leo I could be explained if it were ejected as the lighter member of a bound pair of satellites on its first approach to the Galaxy. These simulations find that this process can generate velocities up to $\sim3$ times the virial velocity (see also Ludlow \etal~2009).

The virial velocity of the group-scale halo surrounding M87 is only $\sim600$ \kms. However, that of Virgo is probably in the range $\sim900$--1300 \kms\ (Strader \etal~2011a), consistent with producing an object like HVGC-1 on the first pericenter passage of an infalling subhalo. However, these extreme velocities are only expected to be observable for a short time after the impulse, so this scenario predicts that HVGC-1 must still be relatively close to the center of M87. The subhalo itself could be observable as a galaxy in the close vicinity of M87.

\subsection{A Hypercompact Stellar System}

Merritt \etal~(2009) and O'Leary \& Loeb (2009) predicted the existence of ``hypercompact" stellar systems in galactic halos, comprised of a SMBH and a population of bound stars. These are the result of asymmetric kicks due to gravitational wave emission during the close interactions of binary BHs. However, the predicted kick velocities are generally lower than observed for HVGC-1, and the internal velocity dispersions are expected to be a significant ($>0.2$) fraction of the kick velocities. Such a large dispersion ($>400$--1000 \kms) far exceeds the upper limit $\la 80$ \kms (\S3.2). Further, HVGC-1 is metal-poor, while hypercompact stellar systems---originating in galaxy centers---should be metal-rich. A related scenario posits the ejection of a hypercompact stellar system in three-body SMBH interactions during multiple galaxy mergers (Kulkarni \& Loeb 2012), but has a similar observational prediction (metal-rich cluster with high $\sigma$), inconsistent with HVGC-1.

\subsection{An Interaction With A Binary Supermassive Black Hole}\label{sec:BH}

When a galaxy with a SMBH is accreted by a central cluster galaxy like M87, the BH will sink to the center via dynamical friction and form a binary SMBH. Stars that pass close enough to this binary can be ejected as hypervelocity stars (Yu \& Tremaine 2003; Holley-Bockelmann \etal~2005). Assuming a 1:10 mass ratio, a primary mass of $6.6\times10^{9} M_{\odot}$ (Gebhardt \etal~2011), and using the Yu (2002) formula, we find that a binary with a separation of $\la 1.7$ pc can eject an object with a velocity $>2300$ \kms. The allowed separation scales with the mass ratio, so separations up to $\sim4.5$ pc are feasible at fixed total mass. 

The same process can apply to GCs, with the important caveat that tidal effects are important. The tidal radius of a $2\times10^{6}M_{\odot}$ GC passing within 1 pc of the M87 SMBH is less than 0.1 pc, so nearly all of the stars would be stripped except for the dense central core. If instead the GC were initially more massive ($\ga10^7M_{\odot}$), and the binary had a 1:3 mass ratio, then the tidal radius would be 0.3--0.4 pc for a distance of 2--3 pc to the BH. If the cluster were relatively dense, most of its stars would still be stripped, but the core, perhaps with $>10^6 M_{\odot}$, could survive and be ejected. Numerical simulations to investigate this possibility are desirable.

We note that Batcheldor \etal~(2010) argued that M87's SMBH is displaced from the stellar center of the galaxy, either because the BH is currently a $\sim$ 1:10 mass ratio binary or because it is undergoing a damped oscillation after a kick from a BH merger. This can be taken as extremely speculative evidence for the recent existence of a binary BH at the center of M87. More concretely, GC kinematics provide evidence for a recent minor merger (Strader \etal~2011a; Romanowsky \etal~2012).

A slight variation on this scenario would be a three-body encounter between a single BH and a \emph{binary} GC, directly analogous to the formation of Galactic hypervelocity stars. Young binary star clusters are known (e.g. Mucciarelli \etal~2012), though these have short coalescence times and no old binary clusters have been discovered.

\subsection{The Future}

While the tangential motion of HVGC-1 is unknown, its radial velocity is so extreme that is it reasonable to assume its tangential motion is smaller than its radial motion. Thus  HVGC-1 is likely to be much further from the center of M87 than the projected distance of $\sim~85$ kpc. If we assume that it originated at the center of M87, then we can calculate its inferred total velocity and compare it to the implied escape velocity for a given halo model.

Under these assumptions, we find that the total velocity of HVGC-1 is easily above that of the escape velocity of Virgo for most published halo models (e.g., Karachentsev \etal~2014;  Rines \& Diaferio 2006), which have virial masses of 4--$8\times10^{14} M_{\odot}.$\footnote{Strader \etal~(2011a) found that no standard NFW (Navarro \etal~1997) halo profile is fully consistent with the kinematic data for M87 and Virgo; nonetheless, HVGC-1 has already left the innermost part of Virgo, so this is less relevant.} HVGC-1 is below escape velocity only in the unlikely case that (a) its 3-D distance is close to its projected distance; (b) the impulse was along our line-of-sight; and (c) Virgo's mass is at the upper end of the allowed range.

Therefore we can conclude that HVGC-1 either will or has escaped from the Virgo Cluster following a three-body interaction---making it the first known hypervelocity GC.

The nature of this interaction remains unclear. A distance to HVGC-1 would help constrain its origin; such a measurement may be possible using deep \emph{Hubble Space Telescope} imaging. At its current motion of $>2.4$ Mpc/Gyr it may have already left the Virgo Cluster and be sailing out into intercluster space.

\acknowledgments

Data from MMT and CFHT; help from SAO/OIR TDC. Support by NSF/AST-1211995 and AST-1109878.

\end{document}